\documentclass[10pt,twocolumn,journal]{IEEEtran}
\usepackage{times}
\usepackage{amsbsy}
\usepackage{latexsym}
\usepackage{amsmath}
\usepackage{color}
\usepackage[ansinew]{inputenc}
\usepackage[dvips ]{graphicx}
\input epsf
\usepackage{epic,eepic}
\usepackage{url}

\IEEEoverridecommandlockouts

\title{\huge Joint Iterative Power Allocation and Linear Interference Suppression
Algorithms in Cooperative DS-CDMA Networks }
\author{Rodrigo C. de Lamare  \\ Communications Research Group \\ Department of Electronics,
    University of York, York Y010 5DD, United Kingdom \\
    Emails: \protect\url{rcdl500@ohm.york.ac.uk}
\thanks{\footnotesize The work of the author was
supported by the University of York, York Y010 5DD, United
Kingdom.   }}

\begin{document}
\maketitle \thispagestyle{empty}

\begin{abstract}

This work presents joint iterative power allocation and interference
suppression algorithms for spread spectrum networks which employ
multiple hops and the amplify-and-forward cooperation strategy for
both the uplink and the downlink. We propose a joint constrained
optimization framework that considers the allocation of power levels
across the relays subject to individual and global power constraints
and the design of linear receivers for interference suppression. We
derive constrained linear minimum mean-squared error (MMSE)
expressions for the parameter vectors that determine the optimal
power levels across the relays and the linear receivers. In order to
solve the proposed optimization problems, we develop cost-effective
algorithms for adaptive joint power allocation, and estimation of
the parameters of the receiver and the channels. An analysis of the
optimization problem is carried out and shows that the problem can
have its convexity enforced by an appropriate choice of the power
constraint parameter, which allows the algorithms to avoid problems
with local minima. A study of the complexity and the requirements
for feedback channels of the proposed algorithms is also included
for completeness. Simulation results show that the proposed
algorithms obtain significant gains in performance and capacity over
existing non-cooperative and cooperative schemes.

\end{abstract}


\section{Introduction}

Multiple collocated antennas enable the exploitation of the spatial
diversity in wireless channels, mitigating the effects of fading and
enhancing the performance of wireless communications systems. Due to
size and cost it is often impractical to equip mobile terminals or
sensor nodes with multiple antennas. However, spatial diversity
gains can be obtained when terminals with single antennas establish
a distributed antenna array through cooperation
\cite{sendonaris}-\cite{laneman04}. In a cooperative transmission
system, terminals or users relay signals to each other in order to
propagate redundant copies of the same signals to the destination
user or terminal. To this end, the designer must employ a
cooperation strategy such as amplify-and-forward (AF)
\cite{laneman04}, decode-and-forward (DF) \cite{laneman04,huang} and
compress-and-forward (CF) \cite{kramer}.

Prior work on cooperative multiuser direct-sequence code-division
multiple-access (DS-CDMA) systems in interference channels has
focused on problems that include the impact of multiple access
interference (MAI) and intersymbol interference (ISI), the problem
of partner selection \cite{huang,venturino} and the bit error rate
(BER) \cite{fang,yang}, outage performance analysis issues
\cite{vardhe}, and the evaluation of the spectral efficiency
\cite{mestre} and the diversity gains \cite{zarifi} based on
asymptotic results. The main motivation for cooperative relaying
with DS-CDMA systems is to increase the capacity, reliability and
the interference suppression capability of these networks
\cite{venturino,vardhe,mestre,zarifi}. Recent contributions in the
area of cooperative communications have considered the problem of
resource allocation \cite{luo,long} in multi-hop time-division
multiple access (TDMA) systems and MIMO systems \cite{tds1,tds2}.
Related work on DS-CDMA system has focused on adaptive modulation
\cite{souryal}, power and rate allocation
\cite{kastrinogiannis,chliu} and scheduling \cite{chen}. In the
literature, there has been no attempt to jointly consider the
problem of power allocation and interference mitigation in
cooperative multiuser DS-CDMA systems so far. This problem is of
paramount importance in cooperative wireless ad-hoc and sensor
networks \cite{souryal}-\cite{kastrinogiannis} that utilize DS-CDMA
systems. These networks require multiple hops to communicate with
nodes that are far from the base station in order to increase their
coverage \cite{jakllari}. Moreover, multi-hop cooperative relaying
can substantially improve the interference suppression capabilities
\cite{venturino,fang,yang}.

{  The goal of this paper is to devise a cross-layer optimization
strategy to significantly increase the capacity, reliability and
coverage of spread spectrum networks which employ multiple hops and
the AF cooperation protocol. Specifically, the problem of joint
resource allocation and linear interference suppression in multiuser
DS-CDMA with a general number of hops is addressed. In order to
facilitate the receiver design, we adopt linear multiuser receivers
\cite{verdu,delamaretvt} which only require a training sequence and
the timing. More sophisticated receiver techniques
\cite{verdu,delamaretc,delamare_itic,stspadf} are also possible for
situations with increased levels of interference. { A joint
constrained optimization framework that considers the allocation of
power levels among the relays subject to individual and global power
constraints and the design of linear receivers is presented. It
should be noted that the proposed design with individual power
constraints has been initially reported in \cite{delamare_vtc},
whereas the proposed design with both individual and global power
constraints has been introduced in \cite{delamare_vtc10}. Here, the
proposed designs are described and investigated in further detail,
more complete derivations along with analysis and simulations
results are included}. Linear MMSE expressions that jointly
determine the optimal power levels across the relays and the linear
receivers are derived. Adaptive least squares (LS) algorithms are
also developed for efficiently solving the joint optimization
problems and mitigating the effects of MAI and ISI, and allocating
the power levels across the links. An analysis of the optimization
problem is conducted and shows that the problem can have its
convexity enforced by an appropriate selection of the power
constraint parameter. This allows the algorithms to avoid problems
with local minima. A study of the computational complexity and the
requirements for feedback channels of the proposed algorithms is
also included.}

{  The main contributions of this work can be summarized as:
\\
1) A joint constrained optimization framework for the allocation of
power levels among the relays subject to individual and global power
constraints and the design of linear receivers; \\ 2) Constrained
linear MMSE expressions for the power allocation and the design of
linear receive filters; \\  3) Recursive algorithms for estimating
the channels, the power allocation and the receive filters;
\\ 4) Convexity analysis of the proposed optimization problems; \\ 5) A
study of the computational complexity and the requirements for
feedback channels of the proposed and existing algorithms.}

{  The rest of this paper is organized as follows. Section II
describes a cooperative DS-CDMA system model with multiple relays.
Section III is devoted to the problem formulation and the
constrained linear MMSE design of the interference mitigation
receiver and the power allocation. The proposed LS algorithms for
the estimation of the receive filter, the power allocation and the
channels subject to a global and individual power constraints are
developed in Sections IV and V, respectively. Section VI is devoted
to the analysis of the computational complexity and feedback
requirements of the proposed algorithms. Section VI presents and
discusses the simulations and Section VII draws the conclusions.}

\section{Cooperative DS-CDMA System and Data Models}

{Consider a synchronous DS-CDMA system communicating over multipath
channels with QPSK modulation, $K$ users, $N$ chips per symbol and
$L$ ($L<N$) as the maximum number of propagation paths for each
link. The synchronous DS-CDMA systems is considered for simplicity
as it captures most of the effects of asynchronous systems with a
low delay spread \cite{delamaretvt,delamaretc}. We consider both
uplink and downlink transmissions. The network is equipped with an
AF protocol that allows communication in multiple hops using $n_r$
relays in a repetitive fashion. {  Therefore, we have $n_p=n_r+1$
phases of transmission or hops and only one transmitter (source or
relay) is active per phase, which increases the delay but also
improves the coverage. The throughput is affected by the fact that
there is an extra time slot per phase of transmission, however,
there are situations for which the performance gains can offset the
extra time slots and the throughput can be improved. } Other
cooperation protocols such as DF can be employed without significant
modifications, however, the AF has been adopted for simplicity and
due to its lower complexity for implementation \cite{laneman04}. We
assume that the source node or terminal transmits data organized in
packets comprising $P$ symbols, where there is a preamble with
training symbols followed by a part with data symbols. We also
assume that the packet contains a sufficient number of training
symbols in the preamble for parameter estimation and that the
network can coordinate transmissions and cooperation. The relays and
destination terminals are equipped with linear receivers, which are
synchronized with their desired signals. Since the focus of this
work is on the resource allocation and linear interference
mitigation, we assume perfect synchronization, however, this
assumption can be relaxed to account for more realistic
synchronization effects in the system. { The proposed algorithms for
power allocation and interference mitigation are employed at the
receivers. A feedback channel is required to convey the power
allocation parameters, which should cope with the channel
variations.} }

\begin{figure}[!htb]
\begin{center}
\def\epsfsize#1#2{1\columnwidth}
\hspace{-0.1em}\epsfbox{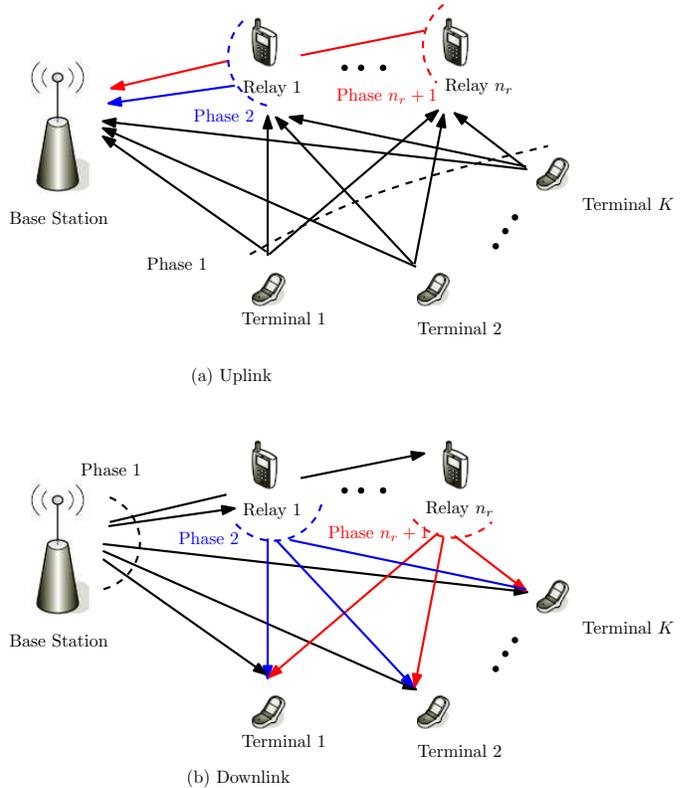} \vspace{-0.5em}\caption{(a) Uplink
and (b) downlink of the cooperative DS-CDMA system.} \label{figsys}
\end{center}
\end{figure}

{The cooperative DS-CDMA system under consideration is depicted in
Fig. \ref{figsys}. The data model is described for the uplink in
what follows. However, it should be remarked that the downlink data
models can be obtained as a particular case of the uplink one. The
received signals are filtered by a matched filter, sampled at chip
rate and organized into $M \times 1$ {vectors ${\boldsymbol
r}_{sd}$, ${\boldsymbol r}_{sr_i}$ and ${\boldsymbol r}_{r_id}$},
which describe the signal received from the source to the
destination, the source to the relays, and the relays to the
destination, respectively,
\begin{equation}
\begin{split}
{\boldsymbol r}_{sd} & = \sum_{k=1}^K  a_{sd}^k {\boldsymbol D}_k
{\boldsymbol h}_{sd,k}b_k  + {\boldsymbol \eta}_{sd} + {\boldsymbol
n}_{sd},
\\
{\boldsymbol r}_{sr_j} & = \sum_{k=1}^K a_{sr_j}^k {\boldsymbol D}_k
{\boldsymbol h}_{sr_j,k} {b}_k  + {\boldsymbol \eta}_{sr_j} +
{\boldsymbol n}_{sr_j},
\\
{\boldsymbol r}_{r_jd} & = \sum_{k=1}^K a_{r_jd}^k {\boldsymbol D}_k
{\boldsymbol h}_{r_jd,k} \tilde{b}_k^{r_jd}  + {\boldsymbol
\eta}_{r_jd} + {\boldsymbol n}_{r_jd}, \label{rvec}
\end{split}
\end{equation} }
{where $M=N+L-1$. The quantity $b_k[m_j]$ corresponds to the
transmitted symbol of user $k$, whereas $\tilde{b}_k^{r_jd}$
represents the symbol $b_k^{r_jd}$ processed at the relay $j$ using
the AF protocol. The amplitudes of the source to the destination,
the source to the relay $j$ and the relay $j$ to the destination
links for user $k$ are denoted by $a_{sd}^k$, $a_{sr_j}^k$ and
$a_{r_jd}^k$, respectively. The vectors ${\boldsymbol n}_{sd}$,
${\boldsymbol n}_{sr_j}$ and ${\boldsymbol n}_{r_jd}$ represent the
noise at the receiver of the destination and the relays. The vectors
${\boldsymbol \eta}_{sd}$, ${\boldsymbol \eta}_{sr_j}$ and
${\boldsymbol \eta}_{r_jd}$ denote the intersymbol interference
(ISI) arising from the source to destination, source to relay $j$
and relay $j$ to destination links, respectively. The $M \times L$
matrix ${\boldsymbol D}_k$ has the signature sequences of each user
shifted down by one position at each column that form }
\begin{equation}
{\boldsymbol D}_k = \left[\begin{array}{c c c }
d_{k}(1) &  & {\bf 0} \\
\vdots & \ddots & d_{k}(1)  \\
d_{k}(N) &  & \vdots \\
{\bf 0} & \ddots & d_{k}(N)  \\
 \end{array}\right],
\end{equation}
{  where ${\boldsymbol d}_k = \big[d_{k}(1), ~d_{k}(2),~ \ldots,~
d_{k}(N) \big]$ stands for the signature sequence of user $k$, the
$L \times 1$ channel vectors from the source to the destination, the
source to the relay, and the relay to the destination are
${\boldsymbol h}_{sd,k}$, ${\boldsymbol h}_{r_jd,k}$, ${\boldsymbol
h}_{r_js,k}$, respectively. By stacking the data vectors in
(\ref{rvec}) (including the links from the relays to the
destination) into a $(n_r+1)M \times 1$ received vector at the
destination we have
\begin{equation}
\begin{split}
\hspace{-0.5em} \left[\hspace{-0.5em}\begin{array}{l}
  {\boldsymbol r}_{sd} \\
  {\boldsymbol r}_{r_{1}d} \\
  \vdots \\
  {\boldsymbol r}_{r_{n_r}d}
\end{array} \hspace{-0.5em} \right] & = \left[\hspace{-0.5em} \begin{array}{l}
  \sum_{k=1}^K  a_{sd}^k {\boldsymbol D}_k {\boldsymbol h}_{sd,k}b_k \\
  \sum_{k=1}^K  a_{{r_1}d}^k {\boldsymbol D}_k {\boldsymbol h}_{{r_1}d,k}{\tilde b}_k^{{r_1}d} \\
  \vdots \\
  \sum_{k=1}^K  a_{{r_{n_r}}d}^k {\boldsymbol D}_k {\boldsymbol h}_{r_{n_r}d,k}{\tilde b}_k^{{r_{n_r}}d}
\end{array}\right] + \left[\hspace{-0.5em} \begin{array}{l}
  {\boldsymbol \eta}_{sd} \\
  {\boldsymbol \eta}_{r_1d} \\
  \vdots \\
  {\boldsymbol \eta}_{r_{n_r}d}
\end{array}\hspace{-0.5em} \right] + \left[ \hspace{-0.5em} \begin{array}{l}
  {\boldsymbol n}_{sd} \\
  {\boldsymbol n}_{r_1 d} \\
  \vdots \\
  {\boldsymbol n}_{r_{n_r}d} \end{array}\right]
\end{split}
\end{equation} }

{  By using the stacked received data from the source and the relays
for joint processing and using $i$ to denote the desired symbol in
the transmitted packet and its received and relayed copies, we can
rewrite the data in a compact form given by}
\begin{equation}
{\boldsymbol r}[i] = \sum_{k=1}^{K} {\boldsymbol {\mathcal C}}_k
{\boldsymbol {\mathcal H}}_k[i] {\boldsymbol B}_k[i] {\boldsymbol
a}_k[i] + {\boldsymbol \eta}[i] + {\boldsymbol n}[i],
\label{recdata}
\end{equation}
where the $(n_r+1)M \times (n_r+1)L$ matrix ${\boldsymbol
{\mathcal C}}_k$ contains shifted versions of ${\boldsymbol D}_k$
as shown by
\begin{equation}
{\boldsymbol {\mathcal C}}_k = \left[\begin{array}{c c c c}
{\boldsymbol D}_{k} & {\bf 0} & \ldots & {\bf 0} \\
{\bf 0} & {\boldsymbol D}_{k} & \ldots & \vdots  \\
\vdots & \vdots & \ddots & {\bf 0} \\
{\bf 0} & {\bf 0} & \ldots &  {\boldsymbol D}_{k}  \\
 \end{array}\right].
\end{equation}

\begin{figure}[!htb]
\begin{center}
\def\epsfsize#1#2{1\columnwidth}
\hspace{-0.1em}\epsfbox{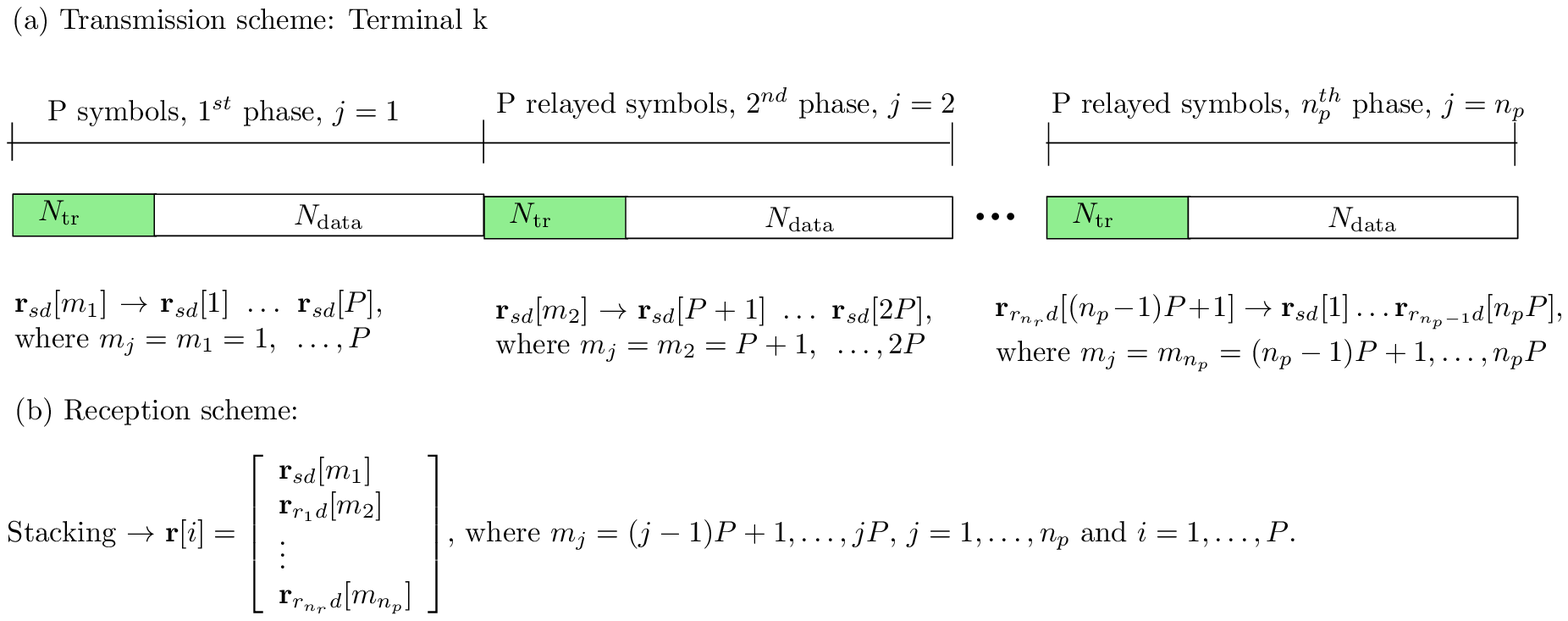} \vspace{-0.5em}\caption{(a)
Transmission and (b) reception scheme of the cooperative DS-CDMA
system.} \label{tscheme}
\end{center}
\end{figure}

{The $(n_r+1)L \times (n_r+1)$ matrix ${\boldsymbol {\mathcal
H}}_k[i]$ has the channel gains of the links between the source and
the destination, and the relays and the destination. { The $(n_r+1)
\times (n_r+1)$ diagonal matrix ${\boldsymbol B}_k[i] = {\rm
diag}(b_k~ {\tilde b}_k^{{r_1}d} \ldots {\tilde b}_k^{{r_n}d}) $
contains the symbols transmitted from the source to the destination
($b_k$) and the $n_r$ symbols transmitted from the relays to the
destination (${\tilde b}_k^{{r_1}d} \ldots {\tilde b}_k^{{r_n}d}$)
on the main diagonal}, the $(n_r+1) \times 1$ vector ${\boldsymbol
a}_k[i]=[a_{sd}^k~a_{{r_1}d}^k\ldots a_{{r_{n_r}}d}^k]^T$ of the
amplitudes, the $(n_r+1)M \times 1$ vector ${\boldsymbol \eta}[i]$
with the ISI terms and $(n_r+1)M \times 1$ vector ${\boldsymbol
n}[i]$ with the noise. A schematic that summarizes the transmission
and reception schemes is depicted in Fig. \ref{tscheme}. }

\section{Problem Statement and Proposed MMSE Design}

This section states the problem of joint power allocation and
interference suppression for a cooperative DS-CDMA network.
Specifically, constrained optimization problems are formulated in
order to describe the joint power allocation and interference
suppression problems subject to a global and individual power
constraints. The proposed linear MMSE designs are aimed at the
destination, which is responsible for jointly computing the receiver
parameters and the power allocation that is sent via a feedback
channel to the source.

\subsection{MMSE Design with a Global Power Constraint}

The linear MMSE design of the power allocation of the links across
the source, relay and destination terminals and interference
suppression filters is presented here using a global power
constraint. Let us express the received vector in (\ref{recdata}) in
a more convenient way for the proposed optimization. The $(n_r+1)M
\times 1$ received vector can be written as
\begin{equation}
{\boldsymbol r}[i] = {\boldsymbol {\mathcal C}}_T {\boldsymbol
{\mathcal H}}_T[i] {\boldsymbol B}_T[i] {\boldsymbol a}_T[i] +
{\boldsymbol \eta}[i] + {\boldsymbol n}[i], \label{recdatat}
\end{equation}
where the $(n_r+1)M \times K(n_r+1)L$ matrix ${\boldsymbol {\mathcal
C}}_T = [ {\boldsymbol {\mathcal C}}_1 ~ {\boldsymbol {\mathcal
C}}_2 ~ \ldots ~  {\boldsymbol {\mathcal C}}_K ]$ contains all the
signatures, the $K(n_r+1)L \times K(n_r+1)$ matrix ${\boldsymbol
{\mathcal H}}_T[i]$ contains the channel gains of {  all the links},
the $K(n_r+1) \times K(n_r+1)$ diagonal matrix ${\boldsymbol B}_T[i]
= {\rm diag}(b_1[m_1]~ {\tilde b}_1^{{r_1}d}[m_2]\ldots {\tilde
b}_1^{{r_{n_r}}d}[m_{n_p}] \ldots b_K[m_1]~ {\tilde
b}_K^{{r_1}d}[m_2] \ldots {\tilde b}_K^{{r_{n_r}}d}[m_{n_p}])$
contains the symbols transmitted from all the sources to the
destination and from all the relays to the destination on the main
diagonal, and the $K(n_r+1) \times 1$ power allocation vector
${\boldsymbol a}_T[i]=[a_{sd}^1[m_1]~a_{{r_1}d}^1[m_2]\ldots
a_{{r_{n_r}}d}^1[m_{n_p}] \ldots a_{sd}^K[m_1]~a_{{r_1}d}^K[m_2]
\ldots a_{{r_{n_r}}d}^K[m_{n_p}]]^T$ contains the amplitudes of all
the links.

Consider {  a joint} MMSE design of the receivers for the $K$ users
represented by a $(n_r+1)M \times K$ parameter matrix ${\boldsymbol
W}[i]=[ {\boldsymbol w}_1[i],~ \ldots, ~ {\boldsymbol w}_K[i]]$ and
for the computation of the $K(n_r +1) \times 1$ optimal power
allocation vector ${\boldsymbol a}_{T,{\rm opt}}[i]$. This problem
can be cast as
\begin{equation}
\begin{split}
[ {\boldsymbol W}_{{\rm opt}}, {\boldsymbol a}_{T,{\rm opt}}  ] &
= \arg \min_{{\boldsymbol W}[i], {\boldsymbol a}_k[i]} ~
E[ ||{\boldsymbol b}[i] - {\boldsymbol W}^H[i]{\boldsymbol r}[i] ||^2 ] \\
{\rm subject ~to~} & {\boldsymbol a}_T^H[i] {\boldsymbol a}_T[i] =
P_{T} , \label{probt}
\end{split}
\end{equation}
where the $K \times 1$ vector ${\boldsymbol b}[i] = [b_1[i], ~
\ldots, b_K[i]]^T$ represents the desired symbols {  of the $K$
users}. The linear MMSE expressions for the parameter matrix
${\boldsymbol W}_{\rm opt}$ and the vector ${\boldsymbol a}_{T, {\rm
opt}}$ can be obtained by transforming the above constrained
optimization problem into an unconstrained one with the method of
Lagrange multipliers \cite{haykin} which leads to
\begin{equation}
\begin{split}
{\mathcal L}_T & = E\big[ \big\|{\boldsymbol b}[i] - {\boldsymbol
W}^H[i]\big({\boldsymbol {\mathcal C}}_T {\boldsymbol {\mathcal
H}}_T[i] {\boldsymbol B}_T[i] {\boldsymbol a}_T[i] +
{\boldsymbol \eta}[i] + {\boldsymbol n}[i]\big) \big\|^2 \big] \\
& \quad + \lambda_T ({\boldsymbol a}_T^H[i] {\boldsymbol a}_T[i]
-P_{T}), \label{lagt}
\end{split}
\end{equation}
Fixing ${\boldsymbol a}_T[i]$, taking the gradient terms of the
Lagrangian and equating them to zero yields
\begin{equation}
{\boldsymbol W}_{{\rm opt}} = {\boldsymbol R}^{-1} {\boldsymbol
P}_{{\boldsymbol{\mathcal C}}{\boldsymbol{\mathcal H}}},
\label{wvect}
\end{equation}
where the covariance matrix of the received vector is ${\boldsymbol
R} = E[{\boldsymbol r}[i]{\boldsymbol r}^H[i]] = {\boldsymbol
{\mathcal C}}_T {\boldsymbol {\mathcal H}}_T[i]{\boldsymbol B}_T[i]
{\boldsymbol a}_T[i] {\boldsymbol a}_T^H[i] {\boldsymbol B}_T^H[i]
{\boldsymbol {\mathcal H}}^H_T[i] {\boldsymbol {\mathcal C}}_T^H +
\sigma^2 {\boldsymbol I}$ and ${\boldsymbol
P}_{{\boldsymbol{\mathcal C}}{\boldsymbol{\mathcal H}}} = E[
{\boldsymbol r}[i]{\boldsymbol b}^H[i]] = E[{\boldsymbol {\mathcal
C}}_T {\boldsymbol {\mathcal H}}_T[i]{\boldsymbol
B}_T[i]{\boldsymbol a}_T[i] {\boldsymbol b}^H[i]] $ is the $(n_r+1)M
\times K$ cross-correlation matrix. The matrices ${\boldsymbol R}$
and ${\boldsymbol P}_{{\boldsymbol{\mathcal C}}{\boldsymbol{\mathcal
H}}}$ depend on the power allocation vector ${\boldsymbol a}_T[i]$.
The expression for ${\boldsymbol a}_T[i]$ is obtained by fixing
${\boldsymbol W}[i]$, taking the gradient terms of the Lagrangian
and equating them to zero which leads to
\begin{equation}
{\boldsymbol a}_{T,{\rm opt}} = ( {\boldsymbol R}_{{\boldsymbol
a}_T} + \lambda_T {\boldsymbol I})^{-1} {\boldsymbol
p}_{{\boldsymbol a}_T}, \label{avect}
\end{equation}
where the $K(n_r+1) \times K(n_r+1)$ covariance matrix ${\boldsymbol
R}_{{\boldsymbol a}_T} = E[{\boldsymbol {\boldsymbol B}}_T^H[i]
{\boldsymbol {\mathcal H}}^H_T[i] {\boldsymbol {\mathcal C}}_T^H
{\boldsymbol W}[i] {\boldsymbol W}^H[i]{\boldsymbol {\mathcal C}}_T
{\boldsymbol {\mathcal H}}_T[i]{\boldsymbol B}_T[i]]$  and the
vector ${\boldsymbol p}_{{\boldsymbol a}_T} = E[{\boldsymbol
B}_T^H[i] {\boldsymbol {\mathcal H}}_T[i]^H {\boldsymbol {\mathcal
C}}_T^H {\boldsymbol W}[i] {\boldsymbol b}[i]]$ is a $K(n_r+1)
\times 1$ cross-correlation vector. The Lagrange multiplier
$\lambda_T$ in the expression above plays the role of a
regularization term and has to be determined numerically due to the
difficulty of evaluating its expression. The expressions in
(\ref{wvect}) and (\ref{avect}) depend on each other and require the
estimation of the channel matrix ${\boldsymbol {\mathcal H}}_T[i]$.
{  Thus, it is necessary to estimate the channel and to iterate
(\ref{wvect}) and (\ref{avect}) with initial values to obtain a
solution.} In addition, the network has to convey all the
information necessary to compute the global power allocation
including the filter ${\boldsymbol W}_{{\rm opt}}$. The expressions
in (\ref{wvect}) and (\ref{avect}) require matrix inversions with
cubic complexity ( $O((n_r+1)M)^3)$ and $O((K(n_r+1))^3)$, should be
iterated as they depend on each other and require channel
estimation.

\subsection{ MMSE Design with Individual Power Constraints}

{  Here, the joint design of a linear MMSE receiver and the
calculation of the optimal power levels across the relays subject to
individual power constraints is presented. Consider an MMSE approach
for the design of the receive filter ${\boldsymbol w}_k[i]$ and the
power allocation vector ${\boldsymbol a}_k[i]$ for user $k$. This
design problem is posed as}
\begin{equation}
\begin{split}
[ {\boldsymbol w}_{k,{\rm opt}}, {\boldsymbol a}_{k,{\rm opt}}  ]
& = \arg \min_{{\boldsymbol w}_k[i], {\boldsymbol a}_k[i]} ~
E[ |b_k[i] - {\boldsymbol w}_k^H[i]{\boldsymbol r}[i] |^2 ] \\
{\rm subject ~to~} & {\boldsymbol a}_k^H[i] {\boldsymbol a}_k[i] =
P_{A,k} , ~~~  k   = 1,~ 2,~\ldots, ~K. \label{prob}
\end{split}
\end{equation}
The expressions for the parameter vectors ${\boldsymbol w}_k[i]$
and ${\boldsymbol a}_k[i]$ can be obtained by transforming the
above constrained optimization problem into an unconstrained one
with the method of Lagrange multipliers \cite{haykin}, which leads
to
\begin{equation}
\begin{split}
{\mathcal L}_k & = E\big[ \big|b_k[i] - {\boldsymbol
w}_k^H[i]\big(\sum_{l=1}^{K} {\boldsymbol {\mathcal C}}_l
{\boldsymbol {\mathcal H}}_l[i] {\boldsymbol B}_l[i] {\boldsymbol
a}_l[i] + {\boldsymbol \eta}[i] + {\boldsymbol n}[i]\big) \big|^2
\big] \\ & \quad + \lambda_k ({\boldsymbol a}_k^H[i] {\boldsymbol
a}_k[i] -P_{A,k}), ~~~  k   = 1,~ 2,~\ldots, ~K. \label{lag}
\end{split}
\end{equation}
Fixing ${\boldsymbol a}_k[i]$, taking the gradient terms of the
Lagrangian and equating them to zero yields
\begin{equation}
{\boldsymbol w}_{k,{\rm opt}} = {\boldsymbol R}^{-1} {\boldsymbol
p}_{{\boldsymbol{\mathcal C}}{\boldsymbol{\mathcal H}}},~~~  k =
1,~ 2,~\ldots, ~K, \label{wvec}
\end{equation}
where ${\boldsymbol R} = \sum_{k=1}^{K}{\boldsymbol {\mathcal
C}}_k {\boldsymbol {\mathcal H}}_k[i]{\boldsymbol B}_k[i]
{\boldsymbol a}_k[i] {\boldsymbol a}_k^H[i] {\boldsymbol B}_k^H[i]
{\boldsymbol {\mathcal H}}^H_k[i] {\boldsymbol {\mathcal C}}_k^H +
\sigma^2 {\boldsymbol I}$ is the covariance matrix and
${\boldsymbol p}_{{\boldsymbol{\mathcal C}}{\boldsymbol{\mathcal
H}}} = E[b_k^*[i] {\boldsymbol r}[i]] = {\boldsymbol {\mathcal
C}}_k {\boldsymbol {\mathcal H}}[i]{\boldsymbol a}_k[i] $ is the
cross-correlation vector. The quantities ${\boldsymbol R}$ and
${\boldsymbol p}_{{\boldsymbol{\mathcal C}}{\boldsymbol{\mathcal
H}}}$ depend on ${\boldsymbol a}_k[i]$. By fixing ${\boldsymbol
w}_k[i]$, the expression for ${\boldsymbol a}_k[i]$ is given by
\begin{equation}
{\boldsymbol a}_{k,{\rm opt}} = ( {\boldsymbol R}_{{\boldsymbol
a}_k} + \lambda_k {\boldsymbol I})^{-1} {\boldsymbol
p}_{{\boldsymbol a}_k}, ~~~  k   = 1,~ 2,~\ldots, ~K, \label{avec}
\end{equation}
where ${\boldsymbol R}_{{\boldsymbol a}_k} = \sum_{k=1}^{K}
{\boldsymbol {\boldsymbol B}}_k^H[i] {\boldsymbol {\mathcal
H}}^H_k[i] {\boldsymbol {\mathcal C}}_k^H {\boldsymbol w}_{k}[i]
{\boldsymbol w}_{k}^H[i]{\boldsymbol {\mathcal C}}_k {\boldsymbol
{\mathcal H}}_k[i]{\boldsymbol B}_k[i]$ is the $(n_r+1) \times
(n_r+1)$ covariance matrix and the $(n_r+1) \times 1$
cross-correlation vector is ${\boldsymbol p}_{{\boldsymbol a}_k} =
E[b_k[i] {\boldsymbol B}_k^H[i] {\boldsymbol {\mathcal H}}_k[i]^H
{\boldsymbol {\mathcal C}}_k^H
{\boldsymbol w}_k[i] ]$. 
{  The expressions in (\ref{wvec}) and (\ref{avec}) have to be
iterated as they depend on each other and require the estimation of
the channel matrices ${\boldsymbol {\mathcal H}}_k[i]$. The
expressions in (\ref{wvec}) and (\ref{avec}) also require matrix
inversions with cubic complexity ( $O(((n_r+1)M)^3)$ and
$O((n_r+1)^3)$.} In what follows, we will develop adaptive
algorithms for computing ${\boldsymbol a}_{k,{\rm opt}}$,
${\boldsymbol w}_{k,{\rm opt}}$ and the channels ${\boldsymbol
{\mathcal H}}_k[i]$ for $k=1, \ldots, K$ in an alternating fashion.

\section{Proposed Joint Estimation Algorithms with a Global Power Constraint}

Here we present adaptive joint estimation algorithms to determine
the parameters of the linear receiver, the power allocation and the
channel with a global power constraint. The proposed joint power
allocation and interference suppression (JPAIS) algorithms with a
global power constraint (GPC) are simply called JPAIS-GPC, are
suitable for the uplink of DS-CDMA systems and rely on LS-based
estimation algorithms. The proposed algorithms are based on the idea
of alternating optimization \cite{csiszar,niesen}, in which the
recursions for computing the parameters of interest are employed in
cycles of iterations and over the received symbols. Note that more
advanced algorithms \cite{delamarespl07}-\cite{barc} could also be
considered in this context.

\subsection{Receiver and Power Allocation Parameter Estimation Algorithms}

Let us now consider the following proposed least squares (LS)
optimization problem
\begin{equation}
\begin{split}
[ \hat{\boldsymbol W}[i], \hat{\boldsymbol a}_{T} [i] ] & = \arg
\min_{{\boldsymbol W}[i], {\boldsymbol a}_T[i]} ~
\sum_{l=1}^{i} \alpha^{i-l} ||{\boldsymbol b}[l] - {\boldsymbol W}^H[i]{\boldsymbol r}[l] ||^2  \\
{\rm subject ~to~} & {\boldsymbol a}_T^H[i] {\boldsymbol a}_T[i] =
P_{T},  \label{probtlst}
\end{split}
\end{equation}
where $\alpha$ is a forgetting factor. The goal is to develop {  a
cost-effective recursive solution} to (\ref{probtlst}). To this end,
we will resort to the theory of adaptive algorithms \cite{haykin}
and derive a constrained joint iterative recursive least squares
(RLS) algorithm. This algorithm will compute $\hat{\boldsymbol
W}[i]$ and $\hat{\boldsymbol a}_T[i]$ and will exchange information
between the recursions. The part of the algorithm to compute
$\hat{\boldsymbol W}[i]$ uses ${\boldsymbol \Phi}[i] =
\hat{\boldsymbol R}[i]= \sum_{l=1}^i \alpha^{l=i} {\boldsymbol
r}[l]{\boldsymbol r}^H[l]$ and is given by
\begin{equation}
{\boldsymbol k}[i] = \frac{\alpha^{-1} {\boldsymbol \Phi}[i]
{\boldsymbol r}[i]}{1+ \alpha^{-1} {\boldsymbol r}^H[i]
{\boldsymbol \Phi}[i] {\boldsymbol r}[i]}, \label{mil1at}
\end{equation}
\begin{equation}
{\boldsymbol \Phi}[i] = \alpha^{-1} {\boldsymbol \Phi}[i-1] -
\alpha^{-1} {\boldsymbol k}[i] {\boldsymbol r}^H[i] {\boldsymbol
\Phi}[i-1], \label{mil1bt}
\end{equation}
\begin{equation} \hat{\boldsymbol W}[i] = \hat{\boldsymbol
W}[i-1] + {\boldsymbol k}[i] {\boldsymbol \xi}^H[i], \label{wrlst}
\end{equation}
where the \textit{a priori} estimation error is given by
\begin{equation}
{\boldsymbol \xi}[i] = {\boldsymbol b}[i] - \hat{\boldsymbol
W}^H[i-1] {\boldsymbol r}[i]. \label{ape1t}
\end{equation}
The derivation for the recursion that estimates the power allocation
$\hat{\boldsymbol a}_T[i]$ presents a difficulty related to the
enforcement of the constraint and how to incorporate it into an
efficient LS algorithm.  This is because the problem in
(\ref{probtlst}) incorporates a Lagrange multiplier ($\lambda_T$) to
ensure the global power constraint and the resulting system of
equations cannot be solved with the aid of the matrix inversion
lemma \cite{haykin} due to its structure. Our approach is to obtain
a recursive expression by relaxing the constraint, to solve the
system of equations with Gaussian elimination or with an adaptive
conjugate gradient algorithm \cite{luen}, and then ensure the
constraint is incorporated via a subsequent normalization procedure.

The power allocation vector ${\boldsymbol a}_T[i]$ is estimated
via the solution of the following equation
\begin{equation}
\hat{\boldsymbol R}_{\hat{\boldsymbol a}_T}[i]\hat{\boldsymbol
a}_T[i] =  \hat{\boldsymbol p}_{\boldsymbol{a}_T}[i],
\label{powalloct}
\end{equation}
where the $K(n_r+1) \times K(n_r+1)$ input data correlation matrix
is given by
\begin{equation}
\begin{split} \hat{\boldsymbol
R}_{\boldsymbol{a}_T}[i] & 
= \alpha \hat{\boldsymbol R}_{\boldsymbol{a}_T}[i-1] +
{\boldsymbol U}_T[i]{\boldsymbol U}_T^H[i],
\end{split}\end{equation} where the $K(n_r+1) \times K$ matrix ${\boldsymbol
U}_T[i] = {\boldsymbol {\boldsymbol B}}_T^H[l] \hat{\boldsymbol
{\mathcal H}}^H_T[l] {\boldsymbol {\mathcal C}}_T^H
\hat{\boldsymbol W}[i]$ and the $K(n_r+1) \times 1$
cross-correlation vector is given by
\begin{equation}
\begin{split}
\hat{\boldsymbol p}_{\boldsymbol{a}_T}[i] & 
= \hat{\boldsymbol
p}_{\boldsymbol{a}_T}[i-1] + {\boldsymbol U}_T[i]{\boldsymbol
b}[i].
\end{split}
\end{equation}
The system of linear equations in (\ref{powalloct}) can be solved
via Gaussian elimination with cubic complexity or more efficiently
with quadratic complexity via an adaptive conjugate gradient
algorithm \cite{luen} as described below
\begin{equation}
{\boldsymbol v}[i] = \hat{\boldsymbol p}_{\boldsymbol{a}_T}[i] -
\hat{\boldsymbol R}_{\hat{\boldsymbol a}_T}[i]\hat{\boldsymbol
a}_T[i], ~~ {\boldsymbol d}[i] = {\boldsymbol v}[i],
\end{equation}
\begin{equation}
\alpha[i] = {\boldsymbol v}^H[i]{\boldsymbol v}[i]/({\boldsymbol
d}^H[i] \hat{\boldsymbol R}_{\hat{\boldsymbol a}_T}[i]{\boldsymbol
d}[i],
\end{equation}
\begin{equation}
\hat{\boldsymbol a}_T[i+1] = \hat{\boldsymbol a}_T[i] + \alpha[i]
\hat{\boldsymbol R}_{\hat{\boldsymbol a}_T}[i]{\boldsymbol d}[i] ,
\label{avec_tableg}
\end{equation}
\begin{equation}
{\boldsymbol v}[i+1] = {\boldsymbol v}[i] - \alpha[i]
\hat{\boldsymbol R}_{\hat{\boldsymbol a}_T}[i]{\boldsymbol d}[i] ,
\end{equation}
\begin{equation}
\beta[i+1] = {\boldsymbol v}^H[i+1]{\boldsymbol
v}[i+1]/({\boldsymbol v}^H[i]{\boldsymbol v}[i]) ,
\end{equation}
\begin{equation}
{\boldsymbol d}[i+1] = {\boldsymbol v}[i+1] + \beta[i+1]
{\boldsymbol d}[i] ,
\end{equation}
In order to ensure the global power constraint $\hat{\boldsymbol
a}_T^H[i]\hat{\boldsymbol a}_T[i] = P_{T}$, we apply the following
rule
\begin{equation}
\hat{\boldsymbol a}_T[i] \leftarrow \sqrt{P_{T}} ~ \hat{\boldsymbol
a}_T[i] \big(\sqrt{\hat{\boldsymbol a}_T^H[i]\hat{\boldsymbol
a}_T[i]}\big)^{-1}. \label{norm_at}
\end{equation}
The algorithms for recursive computation of $\hat{\boldsymbol W}[i]$
and {  $\hat{\boldsymbol a}_T[i]$ } require estimates of the channel
vector ${\boldsymbol {\mathcal H}}_T[i]$, which will also be
developed in what follows. The complexity of the proposed algorithm
is $O(((n_r+1)M)^2)$ for calculating $\hat{\boldsymbol W}[i]$ and
$O((K(n_r+1))^2)$ for obtaining $\hat{\boldsymbol a}_T[i]$.

\subsection{Channel Estimation with a Global Power Constraint}

We present a channel estimator that considers jointly all the $K$
users and exploits the knowledge of the receive filter matrix
$\hat{\boldsymbol W}[i]$ and the power allocation vector
$\hat{\boldsymbol a}_T[i]$. Let us consider the received vector in
(\ref{recdatat}) and develop a channel estimation algorithm for
${\boldsymbol {\mathcal H}}_T[i]$. The proposed channel estimator
can be derived from the following optimization problem
\begin{equation}
\begin{split}
\hat{\boldsymbol {\mathcal H}}_T[i] & = \arg \min_{{\boldsymbol
{\mathcal H}}_T[i]} ~ \sum_{l=1}^{i} \alpha^{i-l} || {\boldsymbol
r}[l] - {\boldsymbol {\mathcal C}}_T {\boldsymbol {\mathcal
H}}_T[i] {\boldsymbol B}_T[i] {\boldsymbol a}_T[i]  ||^2.
\end{split}
\end{equation}
The solution to the above optimization problem is given by
\begin{equation}
\hat{\boldsymbol H}_T[i] = \big( {\boldsymbol {\mathcal C}}_T^H
{\boldsymbol {\mathcal C}}_T \big)^{-1} \hat{\boldsymbol
P}_{{\boldsymbol {\mathcal H}}_T}^{-1}[i] \hat{\boldsymbol
R}_{{\boldsymbol {\mathcal H}}_T}^{-1}[i], \label{cestt}
\end{equation}
where the matrix inversion $\big({\boldsymbol {\mathcal C}}_T^H
{\boldsymbol {\mathcal C}}_T \big)^{-1}$ can be pre-computed and
stored at the terminal of interest, and the $K(n_r+1)L \times
(n_r+1)L$ correlation matrix $\hat{\boldsymbol P}_{{\boldsymbol
{\mathcal H}}_T}[i]$ is computed by the formula
\begin{equation}
\begin{split}
\hat{\boldsymbol P}_{{\boldsymbol {\mathcal H}}_T}[i] & =
\sum_{l=1}^{i} \alpha^{i-l} {\boldsymbol {\mathcal
C}}_T^H{\boldsymbol r}[l] {\boldsymbol u}_{{\boldsymbol {\mathcal
H}}_T}^H[l]\\ & = \hat{\boldsymbol P}_{{\boldsymbol {\mathcal
H}}_T}[i-1] + {\boldsymbol {\mathcal C}}_T^H{\boldsymbol r}[i]
{\boldsymbol u}_{{\boldsymbol {\mathcal H}}_T}^H[i],
\end{split}
\end{equation}
where the $K(n_r+1)L \times 1$ input data vector ${\boldsymbol
u}_{{\boldsymbol {\mathcal H}}_T}[i]$ for this recursion is
\begin{equation}
{\boldsymbol u}_{{\boldsymbol {\mathcal H}}_T}[i] = {{\boldsymbol
B}}_T[i] {\boldsymbol a}_T[i],
\end{equation}
and the inverse of the $K(n_r+1)L \times K(n_r+1)L$ matrix
$\hat{\boldsymbol R}_{{\boldsymbol {\mathcal H}}_T}[i]=
\sum_{l=1}^{i} \lambda^{i-l} {\boldsymbol u}_{{\boldsymbol {\mathcal
H}}_T}[l]{\boldsymbol u}_{{\boldsymbol {\mathcal H}}_T}^H[l]$ is
computed with the aid of the matrix inversion lemma \cite{haykin} as
follows {\small
\begin{equation}
\hat{\boldsymbol R}_{{\boldsymbol {\mathcal H}}_T}^{-1}[i] =
 \frac{\hat{\boldsymbol R}_{{\boldsymbol {\mathcal
H}}_T}^{-1}[i-1]}{\alpha} - \frac{\alpha^{-2} \hat{\boldsymbol
R}_{{h}_k}^{-1}[i-1] {\boldsymbol u}_{{\boldsymbol {\mathcal
H}}_T}[i] {\boldsymbol u}_{{\boldsymbol {\mathcal H}}_T}^H[i]
\hat{\boldsymbol R}_{{\boldsymbol {\mathcal H}}_T}^{-1}[i-1]}{1 +
\alpha^{-1} {\boldsymbol u}_{{\boldsymbol {\mathcal H}}_T}^H[i]
\hat{\boldsymbol R}_{{\boldsymbol {\mathcal H}}_T}^{-1}[i-1]
{\boldsymbol u}_{{\boldsymbol {\mathcal H}}_T}[i]}, \label{milrk}
\end{equation}}
This algorithm jointly estimates the coefficients of the channels
across all the links and for all users subject to a global power
constraint. Therefore, the complexity of the proposed RLS channel
estimation algorithm is $O((K(n_r+1)L)^2)$. { A summary of the main
steps of the GPAIS-GPC algorithms is provided in Table
\ref{summ_gpc}.

\begin{table}[h]
\centering%
\caption{\small Summary of the JPAIS-GPC.} \label{summ_gpc}{
\begin{tabular}{l}
\hline \\
1. Initialise parameters: $\hat{\boldsymbol W}[0]$,
$\hat{\boldsymbol a}_T[0]$ and $\lambda$.\\
for $i=1, \ldots, P$ do\\
2. Compute the receive filter $\hat{\boldsymbol W}[i]$ using
(\ref{wrlst}).\\
3. Calculate the power allocation vector $\hat{\boldsymbol a}_T[i]$
using (\ref{avec_tableg}).\\
4. Normalise $\hat{\boldsymbol a}_T[i]$ with (\ref{norm_at}).\\
5. Compute the channel estimate $\hat{\boldsymbol H}_T[i]$ using
(\ref{cestt}).\\
end for.\\
6. Transmit power allocation $\hat{\boldsymbol a}_T[i]$ to users.\\
 \hline
\end{tabular}
}
\end{table}

}

\section{Proposed Joint Estimation Algorithms with Individual Power
Constraints}

In this section, we present adaptive joint estimation algorithms to
determine the parameters of the linear receiver, the power
allocation and the channel subject to individual power constraints.
The proposed JPAIS algorithms with individual power constraints
(IPC) are simply called JPAIS-IPC and, unlike the algorithms
presented in the previous section, are more appropriate for the
downlink and for distributed resource allocation, detection and
estimation. { Specifically, the JPAIS-IPC algorithms can be employed
in a distributed fashion in which the linear receivers compute the
individual receive filter, the power allocation and channel
parameters, and then convey the power allocation parameters to the
base station.}

\subsection{Receiver and Power Allocation Parameter
Estimation}

We develop a recursive solution to the expressions in (\ref{wvec})
and (\ref{avec}) using time averages instead of the expected
value. The proposed RLS algorithms will compute $\hat{\boldsymbol
w}_k[i]$ and $\hat{\boldsymbol a}_k[i]$ for each user $k$ and will
exchange information between the recursions. We fix
$\hat{\boldsymbol a}_k[i]$ and compute the inverse of
$\hat{\boldsymbol R}[i]$ using the matrix inversion lemma
\cite{haykin} to obtain $\hat{\boldsymbol w}_k[i]$. Defining
${\boldsymbol \Phi}[i] = \hat{\boldsymbol R}[i]$ then we can
obtain the recursions
\begin{equation}
{\boldsymbol k}[i] = \frac{\alpha^{-1} {\boldsymbol \Phi}[i]
{\boldsymbol r}[i]}{1+ \alpha^{-1} {\boldsymbol r}^H[i]
{\boldsymbol \Phi}[i] {\boldsymbol r}[i]}, \label{mil1a}
\end{equation}
\begin{equation}
{\boldsymbol \Phi}[i] = \alpha^{-1} {\boldsymbol \Phi}[i-1] -
\alpha^{-1} {\boldsymbol k}[i] {\boldsymbol r}^H[i] {\boldsymbol
\Phi}[i-1], \label{mil1b}
\end{equation}
\begin{equation}
\hat{\boldsymbol w}_k[i] = \hat{\boldsymbol w}_k[i-1] +
{\boldsymbol k}[i] \xi^*_k[i], \label{wrls}
\end{equation}
where the \textit{a priori} estimation error is
\begin{equation}
\xi_k[i] = b_k[i] - {\boldsymbol w}_k^H[i-1] {\boldsymbol r}[i].
\label{ape1}
\end{equation}
{  The derivation for the recursion that estimates the power
allocation requires a strategy to enforce the constraint on the
individual power of user $k$, which circumvents the need to compute
the Lagrange multiplier $\lambda_k$}. In order to develop the
recursions for $\hat{\boldsymbol a}_k[i]$, we need to compute the
inverse of $\hat{\boldsymbol R}_{\boldsymbol{a}_k}[i]
=\sum_{l=1}^{i} {\boldsymbol {\boldsymbol B}}_k^H[l]
\hat{\boldsymbol {\mathcal H}}^H_k[l] {\boldsymbol {\mathcal C}}_k^H
\hat{\boldsymbol w}_k[l] \hat{\boldsymbol w}^H_k[l]{\boldsymbol
{\mathcal C}}_k \hat{\boldsymbol {\mathcal H}}_k[l]{\boldsymbol
B}_k[l]$. To this end, let us first define ${\boldsymbol
\Phi}_{\boldsymbol {a}_k} = \hat{\boldsymbol
R}_{\boldsymbol{a}_k}[i]$ and proceed as follows:
\begin{equation}
{\boldsymbol k}_{\boldsymbol{a}_k}[i] = \frac{\alpha^{-1}
{\boldsymbol \Phi}_{\boldsymbol {a}_k}[i] {\boldsymbol
B}_k^H[i]\hat{\boldsymbol{\mathcal H}}^H_k[i]{\boldsymbol
{\mathcal C}}_k^H \hat{\boldsymbol w}_k[i]}{1+ \alpha^{-1}
\hat{\boldsymbol w}_k^H[i]{\boldsymbol {\mathcal C}}_k
\hat{\boldsymbol{\mathcal H}}_k[i]{\boldsymbol B}_k[i]
{\boldsymbol \Phi}_{\boldsymbol {a}_k}[i] {\boldsymbol
B}_k^H[i]\hat{\boldsymbol{\mathcal H}}^H_k[i]{\boldsymbol
{\mathcal C}}_k^H \hat{\boldsymbol w}_k[i]}, \label{mil2a}
\end{equation}
\begin{equation}
{\boldsymbol \Phi}_{\boldsymbol {a}_k}[i] = \alpha^{-1}
{\boldsymbol \Phi}_{\boldsymbol {a}_k}[i-1] - \alpha^{-1}
{\boldsymbol k}_{\boldsymbol{a}_k}[i] \hat{\boldsymbol
w}_k^H[i]{\boldsymbol {\mathcal C}}_k \hat{\boldsymbol{\mathcal
H}}_k[i]{\boldsymbol B}_k[i] {\boldsymbol \Phi}_{\boldsymbol
{a}_k}[i-1], \label{mil2b}
\end{equation}
\begin{equation}
\hat{\boldsymbol a}_k[i] = \hat{\boldsymbol a}_k[i] + {\boldsymbol
k}_{\boldsymbol{a}_k}[i] \xi_{\boldsymbol{a}_k}^*[i],
\label{avec_ipc}
\end{equation}
where the \textit{a priori} estimation error for the above
recursion is
\begin{equation}
\xi_{\boldsymbol{a}_k}[i] = b_k[i] - \hat{\boldsymbol
a}_k^H[i]{\boldsymbol B}_k^H[i]\hat{\boldsymbol{\mathcal
H}}^H_k[i]{\boldsymbol {\mathcal C}}_k^H \hat{\boldsymbol w}_k[i],
\end{equation}
In order to ensure the individual power constraint ${\boldsymbol
a}_k^H[i]{\boldsymbol a}_k[i] = P_{A,k}$, we apply the rule
\begin{equation}
\hat{\boldsymbol a}_k[i] \leftarrow \sqrt{P_{A,k}} ~
\hat{\boldsymbol a}_k[i] \big(\sqrt{\hat{\boldsymbol
a}_k^H[i]\hat{\boldsymbol a}_k[i]}\big)^{-1}. \label{norm_ak}
\end{equation}
The algorithms for recursive computation of $\hat{\boldsymbol
w}_k[i]$ and $\hat{\boldsymbol a}_k[i]$ require estimates of the
channel vector ${\boldsymbol {\mathcal H}}_k[i]$, which will also
be developed in what follows. The complexity of the proposed
algorithm is $O(((n_r+1)M)^2)$ for calculating $\hat{\boldsymbol
w}_k[i]$ and $O((n_r+1)^2)$ for obtaining $\hat{\boldsymbol
a}_k[i]$.

\subsection{Channel Estimation with Individual Power Constraints}

We propose here an algorithm that estimates the channels for each
user $k$ across all links subject to individual power constraints
and exploits the knowledge of the receiver filter $\hat{\boldsymbol
w}_k[i]$ and the power allocation vector $\hat{\boldsymbol a}_k[i]$.
Let us consider the received vector in (\ref{recdata}) for the
channel estimation procedure. A channel estimation algorithm can be
developed to solve the following optimization problem
\begin{equation}
\begin{split}
\hat{\boldsymbol {\mathcal H}}_k[i] & = \arg \min_{{\boldsymbol
{\mathcal H}}_k[i]} ~ \sum_{l=1}^{i} \alpha^{i-l} || {\boldsymbol
r}[l] - {\boldsymbol {\mathcal C}}_k{\boldsymbol {\mathcal
H}}_k[i] {\boldsymbol B}_k[l]{\boldsymbol a}_k[l] ||^2,
\\  ~~ {\rm for}~~ k & =1,~2,~\ldots,K,
\end{split}
\end{equation}
The solution to the above optimization problem is
\begin{equation}
\hat{\boldsymbol {\mathcal H}}_k[i] =  \big({\boldsymbol {\mathcal
C}}_k^H {\boldsymbol {\mathcal C}}_k \big)^{-1} \hat{\boldsymbol
P}_{{\boldsymbol h}_k}[i] \hat{\boldsymbol R}_{{\boldsymbol
h}_k}^{-1}[i] , ~~ {\rm for}~~ k=1,~2,~\ldots,K, \label{cestk}
\end{equation}
where the matrix inversion $\big({\boldsymbol {\mathcal C}}_k^H
{\boldsymbol {\mathcal C}}_k \big)^{-1}$ can be pre-computed and
stored at the base station, relays and mobile terminals, and the
$(n_r+1)L \times (n_r+1)L$ correlation matrix $\hat{\boldsymbol
P}_{{\boldsymbol h}_k}[i]$ is computed by the formula
\begin{equation}
\begin{split}
\hat{\boldsymbol P}_{{\boldsymbol h}_k}[i] & = \sum_{l=1}^{i}
\alpha^{i-l} {\boldsymbol {\mathcal C}}_k^H{\boldsymbol r}[l]
{\boldsymbol u}_{{\boldsymbol h}_k}^H[l]\\ & = \hat{\boldsymbol
P}_{{\boldsymbol h}_k}[i-1]+ {\boldsymbol {\mathcal
C}}_k^H{\boldsymbol r}[i] {\boldsymbol u}_{{\boldsymbol
h}_k}^H[i],
\end{split}
\end{equation}
where the $(n_r+1)L \times 1$ input data vector ${\boldsymbol
u}_{{\boldsymbol h}_k}[i]$ for this recursion is
\begin{equation}
{\boldsymbol u}_{{\boldsymbol h}_k}[i] = {{\boldsymbol B}}_k[i]
{\boldsymbol a}_k[i]
\end{equation}
and the inverse of the $(n_r+1)L \times (n_r+1)L$ matrix
$\hat{\boldsymbol R}_{{\boldsymbol h}_k}^{-1}[i]$ is computed as
follows
\begin{equation}
\hat{\boldsymbol R}_{{\boldsymbol h}_k}^{-1}[i] = \alpha^{-1}
\hat{\boldsymbol R}_{{\boldsymbol h}_k}^{-1}[i-1] -
\frac{\alpha^{-2} \hat{\boldsymbol R}_{{h}_k}^{-1}[i-1] {\boldsymbol
u}_{{\boldsymbol h}_k}[i] {\boldsymbol u}_{{\boldsymbol h}_k}^H[i]
\hat{\boldsymbol R}_{{\boldsymbol h}_k}^{-1}[i-1]}{1 + \alpha^{-1}
{\boldsymbol u}_{{\boldsymbol h}_k}^H[i] \hat{\boldsymbol
R}_{{\boldsymbol h}_k}^{-1}[i-1] {\boldsymbol u}_{{\boldsymbol
h}_k}[i]}. \label{milrk}
\end{equation}
This algorithm estimates the coefficients of the channels of each
user $k$ across all the links subject to individual power
constraints. The complexity is $O(((n_r+1)L)^2)$. { A summary of the
main steps of the GPAIS-IPC algorithms is provided in Table
\ref{summ_gpc}.

\begin{table}[h]
\centering%
\caption{\small Summary of the JPAIS-IPC.} \label{summ_ipc}{
\begin{tabular}{l}
\hline \\
1. Initialise parameters: $\hat{\boldsymbol w}_k[0]$,
$\hat{\boldsymbol a}_k[0]$ and $\lambda$.\\
for each user $k$ do \\
for $i=1, \ldots, P$ do\\
2. Compute the receive filter $\hat{\boldsymbol w}_k[i]$ using
(\ref{wrls}).\\
3. Calculate the power allocation vector $\hat{\boldsymbol
a}_k[i]$ using (\ref{avec_ipc}).\\
4. Normalise $\hat{\boldsymbol a}_k[i]$ with (\ref{norm_ak}).\\
5. Compute the channel estimate $\hat{\boldsymbol H}_k[i]$ using
(\ref{cestk}).\\
end for with variable $i$.\\
6. Transmit power allocation $\hat{\boldsymbol a}_k[i]$ to base station.\\

 \hline
\end{tabular}
}
\end{table}

}
\section{Analysis and Requirements of the Algorithms}

In this section, we analyze the optimization problems described in
Section III, and assess the requirements of the proposed JPAIS-GPC
and JPAIS-IPC algorithms in terms of computational complexity and
number of feedback bits.

\subsection{Analysis of the Optimization Problems}

In this part, we develop an analysis of the optimization problems
introduced in Section III, which can be extended to the algorithms
introduced in Sections IV and V. Our approach is based on an
algebraic manipulation of the original problem and the incorporation
of the power constraints to illustrate the properties of the
problem. The optimization problems considered in Section III
constitute non-convex problems, however, it turns out that it is
possible to devise a strategy that modifies the MSE-based
optimization and enforces the convexity. This is corroborated by our
studies that verify that the algorithms converge to the same
solutions regardless of the initialization when this strategy is
adopted.


Let us consider the proposed optimization method in (\ref{probt})
and examine the cost function
\begin{equation}
\begin{split}
{\mathcal{C}}({\boldsymbol W}[i], {\boldsymbol a}_k[i]) & =
E[||{\boldsymbol b}[i] - {\boldsymbol W}^H[i]{\boldsymbol r}[i] ||^2
]\\ & = E[\sum_{k=1}^{K}|{ b}_k[i] - [{\boldsymbol
W}]_k^H{\boldsymbol r}[i] |^2 ], \label{propcfg1}
\end{split}
\end{equation}
which must be minimized subject to ${\boldsymbol a}_T^H[i]
{\boldsymbol a}_T[i] = P_{T}$ and where $[{\boldsymbol W}]_k[i]$
denotes the $k$-column of the matrix ${\boldsymbol W}[i]$. Let us
now express ${\boldsymbol x}[i]= {\boldsymbol W}^H[i]{\boldsymbol
r}[i]$ as
\begin{equation}
\begin{split}
{\boldsymbol x}[i] &  = {\boldsymbol W}^H[i]{\boldsymbol r}[i] \\
&  = {\boldsymbol W}^H[i] (\underbrace{{\boldsymbol {\mathcal
C}}_T {\boldsymbol {\mathcal H}}_T[i] {\boldsymbol
B}_T[i]}_{{\boldsymbol \Re}[i]} {\boldsymbol a}_T[i] +
{\boldsymbol W}^H[i](\underbrace{{\boldsymbol \eta}[i] + {\boldsymbol n}[i]}_{{\boldsymbol t}[i]}) \\
& = \sum_{k=1}^{K}  (\underbrace{[{\boldsymbol W}]_k^H[i]
{\boldsymbol \Re}[i] {\boldsymbol a}_T[i] +  [{\boldsymbol
W}]_k^H[i] {\boldsymbol t}[i]}_{x_k[i]} ) {\boldsymbol \pi}_k,
\end{split}
\end{equation}
where ${\boldsymbol \pi}_k$ is a $K \times 1$ vector that contains a
$1$ in the $k$-th position and zeros elsewhere. Now let us consider
the joint optimization problem via a single parameter vector defined
as
\begin{equation}
{\boldsymbol q}_k[i] = \left[ \begin{array}{c} [{\boldsymbol W}]_k[i] \\
{\boldsymbol a}_T^*[i] \end{array} \right].
\end{equation}
Using the above definition, we can rewrite the data symbol
expression for user $k$ as follows
\begin{equation}
\begin{split}
x_k[i] & = \sum_{k=1}^{K}  \big({\boldsymbol q}_k^H[i] (
{\boldsymbol U}_s[i] + {\boldsymbol U}_I[i]) {\boldsymbol q}_k[i]
\big) {\boldsymbol \pi}_k  \\ & = \sum_{k=1}^{K}  {\boldsymbol
q}_k^H[i] {\boldsymbol U}_T[i]  {\boldsymbol q}_k[i] {\boldsymbol
\pi}_k,
\end{split}
\end{equation}
where the $(n_r+1)(M+K) \times (n_r+1)(M+K)$ matrices with the signal and noise components are \\
${\boldsymbol U}_s[i] = \left[\begin{array}{cc}
{\boldsymbol 0}_{(n_r+1)K \times (n_r+1)K} & {\boldsymbol 0}_{(n_r+1)M \times (n_r+1)M} \\
{\boldsymbol \Re}[i] & {\boldsymbol 0}_{(n_r+1)K \times (n_r+1)K}
\end{array} \right]$ and \\ ${\boldsymbol U}_I[i] = \left[\begin{array}{cc} {\boldsymbol t}[i]
& {\boldsymbol 0}_{(n_r+1)M \times (n_r+1)(K+M)-1}  \\
{\boldsymbol 0}_{(n_r+1)K \times 1} & {\boldsymbol 0}_{(n_r+1)K
\times (n_r+1)(K+M)-1}  \end{array} \right]$.

If we now rewrite the cost function in (\ref{propcfg1}) as a
function of ${\boldsymbol q}_k[i]$, we obtain the equivalent
function
\begin{equation}
{\mathcal{C}}({\boldsymbol q}_k[i])  = E[ \sum_{k=1}^{K}|{ b}_k[i]
- {\boldsymbol q}_k[i]^H{\boldsymbol U}_T[i]{\boldsymbol q}_k[i]
|^2 ], \label{propcfg}
\end{equation}
which is also subject to the global power constraint ${\boldsymbol
a}_T^H[i] {\boldsymbol a}_T[i] = P_{T}$. In order to evaluate the
convexity of the optimization problem, we can verify if the
Hessian (${\boldsymbol H}_k$) with respect to user $k$ of
${\mathcal{C}}({\boldsymbol q}_k[i])$ is positive semi-definite if
${\boldsymbol m}^{H}{\boldsymbol H}_k{\boldsymbol m} \geq 0$ for
all nonzero ${\boldsymbol m} \in
\boldsymbol{C}^{(n_r+1)(M+K)\times 1}$ \cite{luen,golub}.
Computing the Hessian \cite{luen} of the above cost function for
the $k$th user we obtain
\begin{equation}
\begin{split}
{\boldsymbol H}_k & = \frac{\partial}{\partial {\boldsymbol
q}_k^H[i]} \frac{\partial {\mathcal{C}}({\boldsymbol
q}_k[i])}{\partial {\boldsymbol q}_k^H[i]} \\ & = E\big[
({\boldsymbol q}^H_k[i] {\boldsymbol U}_T[i] {\boldsymbol q}_k[i]
-b_k^*[i]) {\boldsymbol U}_T[i] +
 {\boldsymbol U}^H_T[i] {\boldsymbol q}_k[i] {\boldsymbol q}_k^H[i]{\boldsymbol U}_T[i] \\
& \quad  +   ({\boldsymbol q}^H_k[i] {\boldsymbol U}_T[i]
{\boldsymbol q}_k[i] -b_k[i]) {\boldsymbol U}^H_T[i]  +
 {\boldsymbol U}_T[i] {\boldsymbol q}_k[i] {\boldsymbol q}_k^H[i]{\boldsymbol U}^H_T[i]
 \big],
\end{split}
\end{equation}
By examining ${\boldsymbol H}_k$, we verify that the second and
fourth terms are positive semi-definite, whereas the first and the
third terms are indefinite. At this stage, it is of central
importance in the analysis to employ the constraint ${\boldsymbol
a}_T^H[i] {\boldsymbol a}_T[i] = P_{T}$  and perform some
algebraic manipulations in the first and the third terms, which
yield
\begin{equation}
\begin{split}
& 2 \Re e \big\{ E\big[ \big( [{\boldsymbol W}]_k^H[i]
{\boldsymbol \Re}[i] {\boldsymbol \beta}_T[i] P_T  + [{\boldsymbol
W}]_k^H[i]{\boldsymbol t}[i] -b_k^*[i]\big){\boldsymbol U}_T[i]
\big] \big \},
\end{split}
\end{equation}
where ${\boldsymbol \beta}_T[i] = ({\boldsymbol a}_T {\boldsymbol
a}_T^H[i] )^{\dag} {\boldsymbol a}_T$, $(\cdot)^{\dag}$ represents
the pseudo-inverse and the operator $\Re e \{ \cdot \}$ selects the
real part of the argument. The above development shows that the
optimization problem can have its convexity enforced by adjusting
the power constraint $P_T$ so that the following condition holds
\begin{equation}
\begin{split}
P_T  \geq \frac{{\boldsymbol m}^H \Re e \big\{  E\big[
\big(b_k^*[i]- [{\boldsymbol W}]_k^H[i]{\boldsymbol
t}[i]\big){\boldsymbol U}_T[i] \big] \big \} {\boldsymbol m}}{
{\boldsymbol m}^H \Re e \big\{E\big[ \big( [{\boldsymbol
W}]_k^H[i] {\boldsymbol \Re}[i] {\boldsymbol \beta}_T[i]
\big){\boldsymbol U}_T[i]\big] \big \} {\boldsymbol m}}.
\end{split}
\end{equation}
The previous analysis can be conducted for the proposed JPAIS-IPC
algorithm and leads to the following condition
\begin{equation}
\begin{split}
P_k  \geq \frac{{\boldsymbol m}^H \Re e \big\{  E\big[
\big(b_k^*[i] - {\boldsymbol w}_k^H[i]{\boldsymbol j}[i]
\big){\boldsymbol U}_T[i]\big] \big \}{\boldsymbol
m}}{{\boldsymbol m}^H \Re e \big\{  E\big[ \big({\boldsymbol
w}_k^H[i] {\boldsymbol \Re}_k[i] {\boldsymbol
\gamma}_k[i]\big){\boldsymbol U}_T[i] \big] \big \}{\boldsymbol
m}},
\end{split}
\end{equation}
where ${\boldsymbol \Re}_k[i] = {\boldsymbol {\mathcal C}}_k
{\boldsymbol {\mathcal H}}_k[i] {\boldsymbol B}_k[i]$,
${\boldsymbol U}_{k+j}[i] = {\boldsymbol U}_{k}[i] + {\boldsymbol
U}_{j}[i]$, ${\boldsymbol U}_k[i] = \left[\begin{array}{cc}
{\boldsymbol 0}_{(n_r+1) \times (n_r+1)} & {\boldsymbol 0}_{(n_r+1)M \times (n_r+1)M} \\
{\boldsymbol \Re}_k[i] & {\boldsymbol 0}_{(n_r+1) \times (n_r+1)}
\end{array} \right]$ and \\ ${\boldsymbol U}_j[i] = \left[\begin{array}{cc} {\boldsymbol j}[i]
& {\boldsymbol 0}_{(n_r+1)M \times (n_r+1)M-1}  \\
{\boldsymbol 0}_{(n_r+1) \times 1} & {\boldsymbol 0}_{(n_r+1)
\times (n_r+1)M-1}  \end{array} \right]$, the $(n_r+1)M \times 1$
vector ${\boldsymbol j}[i] = \sum_{l\neq k}^K {\boldsymbol
{\mathcal C}}_l {\boldsymbol {\mathcal H}}_l[i] {\boldsymbol
B}_l[i] + {\boldsymbol \eta}[i] + {\boldsymbol n}[i]$, and
${\boldsymbol \gamma}_k[i] = ({\boldsymbol a}_k {\boldsymbol
a}_k^H[i] )^{\dag} {\boldsymbol a}_k$. These conditions have been
verified by numerical experiments, which corroborate the
analytical results. Therefore, the optimization problem can have
its convexity enforced by an appropriate selection of the global
and individual power constraints $P_T$ and $P_k$, respectively.

\subsection{Computational Complexity Requirements}

We discuss here the computational complexity of the proposed and
existing algorithms. Specifically, we will detail the required
number of complex additions and multiplications of the proposed
JPAIS-GPC and JPAIS-IPC algorithms, and compare them with
interference suppression schemes without cooperation (NCIS) and with
cooperation (CIS) \cite{venturino,vardhe} using an equal power
allocation across the relays. Both uplink and downlink scenarios are
considered in the analysis. In Table III we show the computational
complexity required by each recursion associated with a parameter
vector/matrix for the JPAIS-GPC, which is more suitable for the
uplink.

\begin{table}[h]
\centering%
\caption{\small Computational complexity of algorithms with a global
power constraint .} {
\begin{tabular}{ccc}
\hline \rule{0cm}{2.5ex}&  \multicolumn{2}{c}{\small Number of
operations per symbol } \\ \cline{2-3} {\small Parameter} &
{\small Additions} & {\small Multiplications} \\ \hline \\
\emph{ \bf  } & {\small $2((n_r+1)M)^{2} $} & {\small $3((n_r+1)M)^{2}$}   \\
\emph{\bf \small $\hat{\boldsymbol W}[i]$} & {\small $ + 2K((n_r+1)M)$} & {\small $ + 2K((n_r+1)M)$}   \\
\emph{\bf } & {\small $ - (n_r+1)M+1$} & {\small $+ 3(n_r+1)M+1$}
\\ \\\hline \\
\emph{\small \bf }  & {\small $3K(K(n_r+1))$} & {\small $K(K(n_r+1))$} \\
\emph{}  & {\small $+K(n_r+1)(L-1)$} & {\small $+4(K(n_r+1))^2$}
\\
\emph{\small \bf $\hat{\boldsymbol a}_T[i]$ }  & {\small $+K(M(n_r+1))$} & {\small $+(K+L)(K(n_r+1))^2$} \\
\emph{ }  & {\small $+K(K(n_r+1))$} & {\small $-(K(n_r+1))^2$}\\
\emph{}  & {\small $+6(K(n_r+1))^2 $} & {\small $+K(M(n_r+1)L$} \\
\emph{}  & {\small $ + 3K(n_r+1)+n_r+2$} & {\small
$+n_r$} \\
\\\hline \\
\emph{}  & {\small $5(K(n_r+1)L)^{2} $} & {\small $+5(K(n_r+1))^{2} $} \\
\emph{\small \bf $\hat{\boldsymbol H}_T[i]$ }  & {\small $+5K(n_r+1)L$} & {\small $+6K(n_r+1)L$} \\
\emph{}  & {\small $+ 3$} & {\small $+1$} \\ \\ \hline
\end{tabular}
}
\end{table}

\begin{table}[h]
\centering%
\caption{\small Computational complexity of algorithms with
individual power constraints.} {
\begin{tabular}{ccc}
\hline \rule{0cm}{1.5ex}&  \multicolumn{2}{c}{\small Number of
operations per symbol } \\  \cline{2-3} {\small Parameter} &
{\small Additions} & {\small Multiplications} \\ \hline \\
\emph{ \bf  } & {\small $2((n_r+1)M)^{2} $} & {\small $3((n_r+1)M)^{2}$}   \\
\emph{\bf \small $\hat{\boldsymbol w}_k[i]$} & {\small $ + (n_r+1)M$} & {\small $ + 5(n_r+1)M$}   \\
\emph{\bf } & {\small $ +1$} & {\small $+1$}
\\ \\\hline \\
\emph{\small \bf }  & {\small $2(n_r+1)^2$} & {\small $3(n_r+1)^2$} \\
\emph{}  & {\small $+3(n_r+1)$} & {\small $+7(n_r+1)$} \\
\emph{\small \bf $\hat{\boldsymbol a}_k[i]$ }  & {\small $+M(n_r+1)L$} & {\small $+M(n_r+1)L$} \\
\emph{ }  & {\small $+(n_r+1)L$} & {\small $+(n_r+1)L$}\\
\emph{}  & {\small $-3 $} & {\small
$+3$} \\
\\\hline \\
\emph{}  & {\small $2((n_r+1)L)^{2} $} & {\small $6((n_r+1)L)^{2} $} \\
\emph{\small \bf $\hat{\boldsymbol H}_k[i]$ }  & {\small $+5M(n_r+1)L$} & {\small $+M(n_r+1)L$} \\
\emph{}  & {\small $-5(n_r+1)+ 3$} & {\small $+4(n_r+1)+1$} \\ \\
\hline
\end{tabular}
}
\end{table}

In Table IV we describe the computational complexity required by
each recursion associated with a parameter vector for the JPAIS-IPC
algorithm, which is suitable for both the uplink and the downlink. A
noticeable difference between the JPAIS-GPC and the JPAIS-IPC is
that the latter is employed for each user, whereas the former is
used for all the $K$ users in the system. Since the computation of
the inverse of $\hat{\boldsymbol R}[i]$ is common to all users for
the uplink in our system, the JPAIS-GPC is more efficient than the
JPAIS-IPC when the latter is computed for all the $K$ users.

\begin{table}[h]
\centering%
\caption{\small Computational complexity of the proposed JPAIS and
existing algorithms.} {
\begin{tabular}{ll}
\hline \rule{0cm}{2.5ex} {\small Algorithm} &
{\small Recursions }  \\ \hline \\
\emph{ \bf  JPAIS-GPC (Uplink) } & {\small $\hat{\boldsymbol W}[i]$, $\hat{\boldsymbol a}_T[i]$,$\hat{\boldsymbol H}_T[i]$}   \\
 \\\hline \\
\emph{\small \bf JPAIS-IPC (Downlink) }  & {\small $\hat{\boldsymbol w}_k[i]$, $\hat{\boldsymbol a}_k[i]$, $\hat{\boldsymbol H}_k[i]$ }  \\
\\\hline \\
\emph{\small \bf CIS (Uplink) }  & {\small $\hat{\boldsymbol W}[i]$, $\hat{\boldsymbol a}_T[i]$ is fixed}  \\
\\ \hline \\
\emph{\small \bf CIS (Downlink) }  & {\small $\hat{\boldsymbol w}_k[i]$, $\hat{\boldsymbol a}_k[i]$ is fixed}  \\
\\ \hline  \\
\emph{\small \bf NCIS (Uplink)}  & {\small $\hat{\boldsymbol W}[i]$ with $n_r=0$}  \\
\\ \hline \\
\emph{\small \bf NCIS (Downlink)}  & {\small $\hat{\boldsymbol w}_k[i]$ with $n_r=0$}  \\
\\ \hline

\end{tabular}
}
\end{table}

The recursions employed for the proposed JPAIS-GPC and the JPAIS-IPC
are general and parts of them are used in the existing CIS and NIS
algorithms. Therefore, we can use them to describe the required
computational complexity of the existing algorithms. In Table V we
show the required recursions for the proposed and existing
algorithms, whose complexity is detailed in Tables III and IV.

\begin{figure}[!htb]
\begin{center}
\def\epsfsize#1#2{1\columnwidth}
\epsfbox{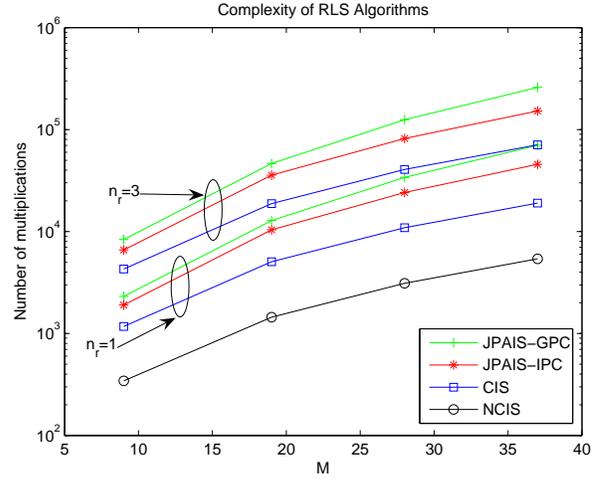} \vspace{-1.0em}\caption{\footnotesize
Computational complexity in terms of the number of complex
multiplications of the proposed and existing schemes for the
uplink.} \vspace{-0.75em}\label{fig_comp}
\end{center}
\end{figure}

In Fig. \ref{fig_comp}, we illustrate the required computational
complexity for the proposed and existing schemes for different
number of relays ($n_r$). The curves show that the proposed
JPAIS-GPC and JPAIS-IPC are more complex than the CIS scheme and the
NCIS.  This is due to the fact that the power allocation and channel
estimation recursions are employed. {  However, we will show in the
next section that this additional required complexity (which is
modest and equivalent to an additional $20-25 \%$ cost as compared
to the CIS scheme) can significantly improve the performance of the
system.}

\subsection{Feedback Channel Requirements}

\begin{figure}[!htb]
\begin{center}
\def\epsfsize#1#2{1\columnwidth}
\epsfbox{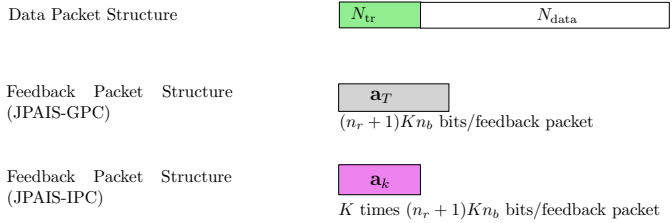} \vspace{-0.25em} \caption{\footnotesize Proposed
structure of the data and feedback packets.} \label{fig_packet}
\end{center}
\end{figure}

The proposed JPAIS algorithms require feedback signalling in order
to allocate the power levels across the relays. In order to
illustrate how these requirements are addressed, we can refer to
Fig. \ref{fig_packet} which depicts the structure for both the data
and feedback packets. {  The data packet comprises a preamble with a
number of training symbols ($N_{\rm tr}$), which are used for
parameter estimation and synchronization, and the transmitted data
symbols ($N_{\rm data}$)}. The feedback packet requires the
transmission of the power allocation vector ${\boldsymbol a}_T$ for
the case of the JPAIS-GPC algorithm, whereas it requires the
transmission of ${\boldsymbol a}_k$ for each user for JPAIS-IPC. A
typical number of bits $n_b$ required to quantize each coefficient
of the vectors ${\boldsymbol a}_T$ and ${\boldsymbol a}_k$ via
scalar quantization is $n_b=4$ bits. More efficient schemes
employing vector quantization \cite{gersho,delamare_ieeproc} and
that take into account correlations between the coefficients are
also possible.

For the uplink (or multiple-access channel), the base station (or
access point) needs to feedback the power levels across the links to
the $K$ destination users in the system. With the JPAIS-GPC
algorithm, the parameter vector ${\boldsymbol a}_T$ with $(n_r+1)K
n_b$ bits/packet must be broadcasted to the $K$ users. For the
JPAIS-IPC algorithm, a parameter vector ${\boldsymbol a}_k$ with
$(n_r+1) n_b$ bits/packet must be broadcasted to each user in the
system. In terms of feedback, the JPAIS-IPC algorithm is more
flexible and may require less feedback bits if there is no need for
a constant update of the power levels for all $K$ users.

For the downlink (or broadcast channel), the $K$ users must
feedback the power levels across the links to the base station.
With the JPAIS-GPC algorithm, the parameter vector ${\boldsymbol
a}_T$ with $(n_r+1)K n_b$ bits/packet must be computed by each
user and transmitted to the base station, which uses the
${\boldsymbol a}_T$ vector coming from each user. An algorithm for
data fusion or a simple averaging procedure can be used. For the
JPAIS-IPC algorithm, a parameter vector ${\boldsymbol a}_k$ with
$(n_r+1) n_b$ bits/packet must be transmitted from each user to
the base station. In terms of feedback, the JPAIS-IPC algorithm
requires significantly less feedback bits than the JPAIS-GPC in
this scenario.

\section{Simulations}

{ In this section, a simulation study of the proposed JPAIS and
existing algorithms is carried out. The first existing scheme that
is considered in the comparisons is a linear interference
suppression technique that only takes into account the source to
destination links and does not consider the contribution of the
relays. This scheme is denoted non-cooperative interference
suppression (NCIS) and corresponds to a linear receive filter
designed according to the MMSE criterion or computed with an RLS
algorithm for each user. The second existing scheme is denoted
cooperative interference suppression (CIS) and processes the signals
arriving from the source and the relays using an equal power
allocation across the relays for each user. The CIS scheme employs a
linear receive filter designed according to the MMSE criterion or
adjusted with an RLS algorithm, and the entries of the power
allocation parameter vectors are equal (equal power allocation). We
first evaluate the bit error ratio (BER) performance of the proposed
JPAIS-GPC and JPAIS-IPC algorithms and compare them with the NCIS
and the CIS schemes.} We consider a DS-CDMA system with randomly
generated spreading codes with a processing gain $N=16$. {The noise
samples at the receivers of the relays and the destination are drawn
from zero mean complex Gaussian random variables with variance
$\sigma^2$}. The fading channels (that can be time-varying or
time-invariant) are generated using a random power delay profile
with gains taken from a complex Gaussian variable with unit variance
and mean zero, $L=3$ paths spaced by one chip, and are normalized
for unit power. The time-varying channels are generated according to
Clarke's model \cite{rappa}, which is parameterized by the
normalized Doppler frequency $f_d T$, where $f_d$ is the Doppler
frequency and $T$ is the inverse of the symbol rate. The power
constraint parameter $P_{A,k}$ is set for each user so that one can
control the SNR (${\rm SNR} = P_{A,k}/\sigma^2$) and $P_T=K
P_{A,k}$, whereas the power distribution of the interferers follows
a log-normal distribution with associated standard deviation of $3$
dB. We adopt the AF cooperative strategy with repetitions and all
the relays and the destination terminal are equipped with linear
MMSE or adaptive receivers. Note that the noise amplification of the
AF protocol is considered \cite{laneman04}. The receivers have
either full knowledge of the channel and the noise variance ( MMSE
design) or are adaptive and estimate all the required coefficients
and the channels using the proposed and existing algorithms with
optimized parameters. {  For the JPAIS algorithms employing the MMSE
expressions we employ $2$ iterations per packet (when the channels
are time-invariant) or per symbol (when the channels are
time-varying) for the design of the parameter vectors, whereas for
the adaptive versions we use only $1$ iteration per update.} The
JPAIS algorithms are used at the destination and employ a feedback
channel to send the power allocation vector to the source, whereas
the relays are equipped with conventional linear MMSE or adaptive
receivers. We employ packets with $1500$ QPSK symbols and average
the curves over $1000$ runs. For the adaptive receivers, we provide
training sequences with $N_{\rm tr}=200$ symbols placed at the
preamble of the packets. After the training sequence, the adaptive
receivers are switched to decision-directed mode.

\begin{figure}[!htb]
\begin{center}
\def\epsfsize#1#2{1\columnwidth}
\epsfbox{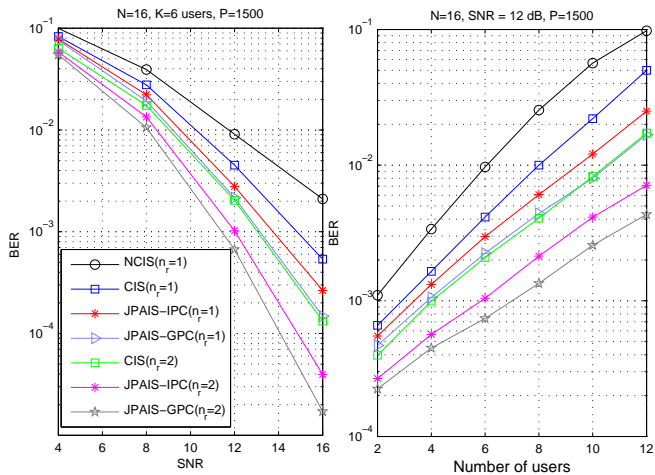} \vspace{-1.25em} \caption{\footnotesize BER
versus SNR and $K$ for the optimal linear MMSE detectors.
Parameters: $\lambda_T = \lambda_k =0.025$.}
\vspace{-0.75em}\label{fig1}
\end{center}
\end{figure}

We first consider the proposed JPAIS method with the MMSE
expressions of (\ref{wvect}) and (\ref{avect}) using a global power
constraint (JPAIS-GPC), and (\ref{wvec}) and (\ref{avec}) with
individual power constraints (JPAIS-IPC). We compare the proposed
scheme with a non-cooperative approach (NCIS) and a cooperative
scheme with equal power allocation (CIS) across the relays for
$n_r=1,2$ relays. The results shown in Fig. \ref{fig1} illustrate
the performance improvement achieved by the proposed JPAIS scheme
and algorithms, which significantly outperform the CIS and the NCIS
techniques. As the number of relays is increased so is the
performance, reflecting the exploitation of the spatial diversity.
In the scenario studied, the proposed JPAIS-IPC approach can
accommodate up to $3$ more users as compared to the CIS scheme and
double the capacity as compared with the NCIS for the same
performance. The proposed JPAIS-GPC is superior to the JPAIS-IPC and
can accommodate up to $2$ more users than the JPAIS-IPC, while its
complexity is higher. Equivalently, the BER versus SNR curves show
that the proposed JPAIS scheme and algorithms can obtain a higher
diversity order than the existing schemes, and save up to $2$ dB in
SNR for the same BER as compared with the existing techniques. { The
results in Fig.  \ref{fig1} suggest that the JPAIS-GPC algorithms
are more suitable than the JPAIS-IPC algorithms for the uplink and
situations with a high SNR and a large number of users. For the
downlink and situations with low SNR and a small number of users,
the gains of the JPAIS-GPC algorithms over the JPAIS-IPC algorithms
are more very significant, suggesting that the latter are more
suitable in these scenarios.} {The reason for the improved
performance of the JPAIS algorithms over the existing schemes is
that they jointly optimize the linear receive filter parameters and
the power allocation, better exploiting the degrees of freedom at
both the transmitter via power allocation and at the receiver with
linear interference suppression to mitigate the interference. This
approach allows a more effective reduction of the MSE and an
improvement in the BER.}

\begin{figure}[!htb]
\begin{center}
\def\epsfsize#1#2{1\columnwidth}
\epsfbox{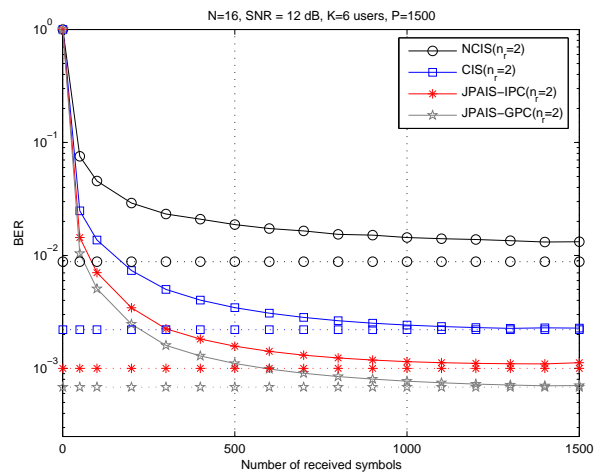} \vspace{-1.0em}\caption{\footnotesize BER
performance versus number of symbols. The curves for the adaptive
schemes are in solid lines, whereas those of the optimal MMSE
schemes are in dotted lines. Parameters: $\lambda_T=\lambda_k=0.025$
(for MMSE schemes), $\alpha=0.998$ (for adaptive schemes).}
\vspace{-0.75em}\label{fig3}
\end{center}
\end{figure}

The second experiment depicted in Fig. \ref{fig3} shows the BER
performance of the proposed adaptive algorithms (JPAIS) against the
existing NCIS and CIS schemes with $n_r=2$ relays. All techniques
employ RLS algorithms for the estimation of the coefficients of the
channel, the receive filters and the power allocation for each user
(JPAIS only). The complexity of the proposed algorithms is quadratic
with the filter length of the receivers and the number of relays
$n_r$, whereas the optimal MMSE schemes require cubic complexity. {
From the results, we can verify that the proposed adaptive joint
estimation algorithms converge to approximately the same level of
the MMSE schemes, which have full knowledge of the channel and the
noise variance.} Again, the proposed JPAIS-GPC is superior to the
JPAIS-IPC but requires higher complexity and joint demodulation of
signals, whereas the JPAIS-IPC lends itself to a distributed
implementation.

\begin{figure}[!htb]
\begin{center}
\def\epsfsize#1#2{1\columnwidth}
\epsfbox{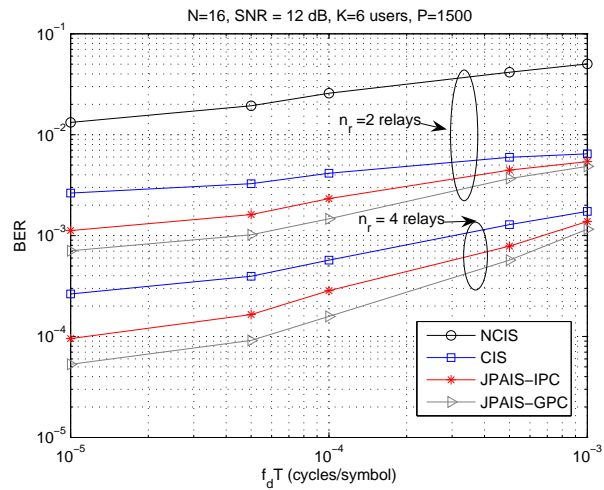} \vspace{-1.0em}\caption{\footnotesize BER
performance versus $f_dT$. The parameters of the adaptive algorithms
are optimized for each $f_dT$.} \vspace{-0.15em}\label{fig6}
\end{center}
\end{figure}

The next experiment considers the average BER performance against
the normalized fading rate $f_dT$ (cycles/symbol), as depicted in
Fig. \ref{fig6}. The idea is to illustrate a situation where the
channel changes within a packet and the system transmits the power
allocation vectors computed by the proposed JPAIS algorithms via a
feedback channel. In this scenario, the JPAIS algorithms compute the
parameters of the receiver and the power allocation vector, which is
transmitted only once to the mobile users. This leads to a situation
in which the power allocation becomes outdated. The results show
that the gains of the proposed JPAIS algorithms decrease gradually
as the $f_dT$ is increased to the BER level of the existing CIS
algorithms for both $n_r=2$ and $n_r=4$ relays, indicating that the
power allocation is no longer able to provide performance
advantages. This problem requires the deployment of a frequent
update of the power allocation via feedback channels. { Therefore,
the proposed JPAIS algorithms are suitable for scenarios with $f_dT
< 10^{-3}$ for which there are performance gains.}

\begin{figure}[!htb]
\begin{center}
\def\epsfsize#1#2{1\columnwidth}
\epsfbox{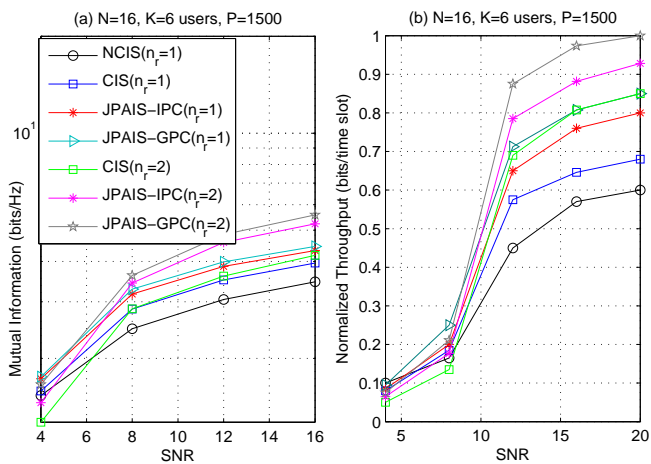} \vspace{-1.0em}\caption{\footnotesize (a) Mutual
information (bits/Hz) versus SNR (dB) and (b) the normalized
throughput (bits/time slot) versus SNR (dB).}
\vspace{-0.15em}\label{newfig}
\end{center}
\end{figure}

{The algorithms are now assessed in terms of the mutual information
between the $k$-th user and the base station as suggested in
\cite{vardhe} and the normalized throughput (NT) defined as $NT =
R(1 - BER)^{P ~log_2 M}$ in bits/time slot, where $R$ is the
normalized rate, $M$ is the number of points in the constellation
and $P$ is the packet size in symbols. Since the protocols operate
at full rate in a synchronous system and QPSK modulation is used the
parameter used are $R=1$ and $M=4$. The results are shown in Fig.
\ref{newfig} and indicate that the JPAIS algorithms can obtain gains
in terms of the mutual information and the NT for a sufficiently
high SNR level. When the SNR is low the use of more phases of
transmission (or time slots) can degrade the NT and the mutual
information of the system. The use of the proposed JPAIS algorithms
with multihop transmission is not recommended in these situations.
However, as the SNR is increased
 the proposed JPAIS algorithms obtain the best
results with $n_r = 1, 2$. The JPAIS-GPC algorithm achieves the best
NT results followed by the JPAIS-IPC and the CIS techniques.}

\begin{figure}[!htb]
\begin{center}
\def\epsfsize#1#2{1\columnwidth}
\epsfbox{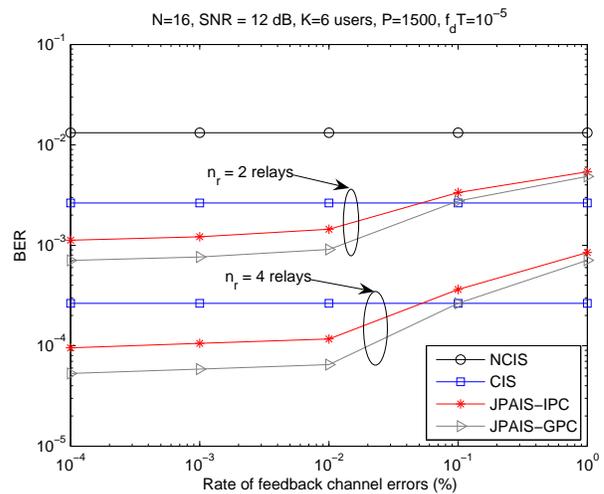} \vspace{-1.0em}\caption{\footnotesize BER
performance versus $f_dT$. The parameters of the adaptive algorithms
are optimized for each $f_dT$.} \vspace{-0.25em}\label{fig7}
\end{center}
\end{figure}

The last experiment, shown in Fig. \ref{fig7}, illustrates the
averaged BER performance versus the percentage of errors in the
feedback channel for an uplink scenario. Specifically, the feedback
packet structure is employed and each coefficient is quantized with
$4$ bits. Each feedback packet is constructed with a sequence of
binary data ($0$s and $1$s) and is transmitted over a binary
symmetric channel (BSC) with an associated probability of error
$P_e$. We then evaluate the BER of the proposed JPAIS and the
existing algorithms against several values of the $P_e$. The results
show that the proposed JPAIS algorithms obtain significant gains
over the existing CIS algorithm for values of $P_e < 0.1 \%$. {  As
we increase the rate of feedback errors, the performance of the
proposed JPAIS algorithms becomes worse than the CIS algorithms and
are no longer suitable.} This suggests the use of error-control
coding techniques to keep the level of errors in the feedback
channel below a certain value.

\section{Concluding remarks and extensions}

We have presented in this work joint iterative power allocation and
interference mitigation techniques for DS-CDMA networks which employ
multiple hops and the AF cooperation strategy. A joint constrained
optimization framework and algorithms that consider the allocation
of power levels across the relays subject to global and individual
power constraints and the design of linear receivers for
interference suppression have been proposed. A study of the proposed
optimization problems has been carried out and has shown that the
convexity of the problem can be enforced via an appropriate choice
of the global and individual power constraints. A study of the
requirements of the proposed and existing algorithms in terms of
computational complexity and feedback channels has also been
conducted. The results of simulations have shown that the proposed
JPAIS techniques obtain significant gains in performance and
capacity over existing non-cooperative and cooperative schemes. The
proposed JPAIS algorithms can be employed in a variety of wireless
communications systems with relays including multiple-antenna,
orthogonal-frequency-division-multiplexing (OFDM) and ultra-wide
band (UWB) systems. {  Prior work on asymptotic results has been
reported in \cite{mestre} and \cite{zarifi} for $2$-phase systems
without power allocation. A possible extension would be to use the
power allocation strategy of this work and investigate the
asymptotic gains. }

\end{document}